\UseRawInputEncoding
\documentclass[aps,prd,preprintnumbers,groupedaddress,superscriptaddress,floatfix,tightenlines,twocolumn,reprint,nofootinbib]{revtex4-2}
\usepackage{units}
\usepackage{graphicx}
\usepackage{dcolumn}
\usepackage{bm}
\usepackage{leftidx}
\usepackage{bbold}
\usepackage{amssymb}
\usepackage{amsfonts,amsmath,epsfig,bm,physics}
\usepackage[dvipsnames]{xcolor}
\usepackage{graphicx,hyperref}
\usepackage{booktabs}
\usepackage{esvect}
\usepackage[caption=false]{subfig}
\hypersetup{
    colorlinks,
    citecolor=blue,
    filecolor=red,
    linkcolor=Maroon,
    urlcolor=violet
}
\usepackage{mathtools}          

\begin{document}


\title{Primordial black hole formation in $\alpha$-attractor models: an analysis using optimized peaks theory}

\author{Rafid Mahbub}
\email{mahbu004@umn.edu}

\affiliation{%
 School of Physics \& Astronomy, University of Minnesota,\\
 Minneapolis, MN 55455, USA 
}%

\begin{abstract}
    In this paper, the formation of primordial black holes (PBHs) is reinvestigated using inflationary $\alpha$-attractors. Instead of using the conventional Press-Schechter theory to compute the abundance, the optimized peaks theory is used, which was developed in Ref. \cite{Yoo:2018kvb,Yoo:2020dkz}. This method takes into account how curvature perturbations play a r\^{o}le in modifying the mass of primordial black holes. Analyzing the model proposed in \cite{Mahbub:2019uhl} it is seen that the horizon mass of the collapsed Hubble patch is larger by $\mathcal{O}(10)$ compared to the usual computation. Moreover, PBHs can be formed from curvature power spectrum, $\mathcal{P}_{\zeta}(k)$, peaked at lower values using numerically favored threshold overdensities. As a result of the generally larger masses predicted, the peak of the power spectrum can be placed at larger $k$ modes than that is typical with which potential future constraints on the primordial power spectrum through gravitational waves (GWs) can be evaded.
\end{abstract}

\maketitle

\section{Introduction}
Primordial black holes are thought to have been formed in the early Universe from the collapse of rare and large overdense regions \cite{Carr:1974nx,Carr:1975qj,Zeldovich:1967lct}. The density fluctuations arise from inflation when the quantum fluctuations of the inflaton field get stretched to superhorizon scales and later reenter the horizon during the post-inflationary Universe \cite{1979ZhPmR..30..719S,Guth:1980zm,Linde:1983gd,Mukhanov:1981xt}. The density fluctuations that could give rise to PBHs become subhorizon during the radiation dominated (RD) epoch and need to be large in order for gravitational collapse to take place. In general, these need to be $\delta\rho/\rho \sim 10^{-1}$. However, inflation models usually predict this to be $\delta\rho/\rho\sim 10^{-5}$ in a nearly scale-invariant fashion and such small overdensities cannot give rise to a cosmologically interesting number of PBHs- unless the primordial curvature power spectrum $\mathcal{P}_{\zeta}(k)$ is parameterized by a blue-tilted scalar spectral index $n_{s}$ \cite{Young:2014ana,Green:2018akb}. As a result of this, people have increasingly looked towards inflation models which possess certain features which can give rise to amplified overdensity perturbations at small scales \cite{Ozsoy:2018flq,Ozsoy:2020kat,Dalianis:2018frf,Mishra:2019pzq}. These amplified overdensities can then collapse to form a cosmologically significant population of PBHs which could then explain the nature of cold dark matter (CDM) \cite{Carr:2020xqk,Carr:2020gox}.\\
\indent The abundance of PBHs is usually calculated using the Press-Schechter (PS) formalism \cite{1974ApJ...187..425P}. But the PS formalism requires the smoothing of $\delta(t,\bm{x})$ over some lengthscale, $R$, using a suitably defined window function.\footnote{Smoothing removes structure on scales $\lesssim R$. This then ensures that large fluctuations on small scales do not contribute to structure formation at some larger scale.} This is needed in order to induce a scale dependence (and hence, a mass dependence) on the abundance of virialized objects formed from gravitational collapse. A problem arises from the fact that, although there are various window functions, there is no good way of ascertaining which one is appropriate for studying PBHs and different choices lead to very different outcomes \cite{Ando:2018qdb,Young:2019osy}. Another approach is to use peaks theory (which computes the number of peaks in the overdensity field) \cite{1986ApJ...304...15B} which has seen an increased use in the study of PBHs. Being developed under a more rigorous mathematical framework, peaks theory also enables one to incorporate the shape dependence of the profile of curvature perturbations and not simply establishing a collapse criterion using a threshold $\delta_{\text{th}}$. Of particular importance is the fact that, in peaks theory, smoothing is not necessary since the scale dependence of the PBH fraction can be induced naturally without the need of smoothing. Another important reason why the shape of perturbation profiles should be incorporated into the study of PBHs is in regards to mass calculations. Usually one associates the mass of the collapse Hubble patch to that of the PBH formed. However, it is common to use the horizon crossing condition in Fourier space $k=a(t)H(t)$, which is useful only when the Hubble patch is described by small perturbations. In fact, larger perturbations (of the order of interest in PBH formation) lead to an order of magnitude error in the horizon mass calculated using the na\"{i}ve approach. The optimized peaks formalism developed in \cite{Yoo:2018kvb} addresses these issues. Furthermore, the optimized peaks calculation could potentially allow PBH formation to be studied using smaller amplitudes of $\mathcal{P}_{\zeta}(k)$. Although there are no stringent bounds on the size of the primordial power spectrum at PBH forming scales, future constraints from the Laser Interferometer Gravitational-Wave Observatory (LISA) and the Square Kilometer Array (SKA) can push this to lower values \cite{Byrnes:2018txb,Inomata:2018epa,Green:2020jor}.

\section{Inflationary $\alpha$-attractor model}\label{sec:sectionII}
\begin{figure}[h]
\centering
\includegraphics[scale=0.6]{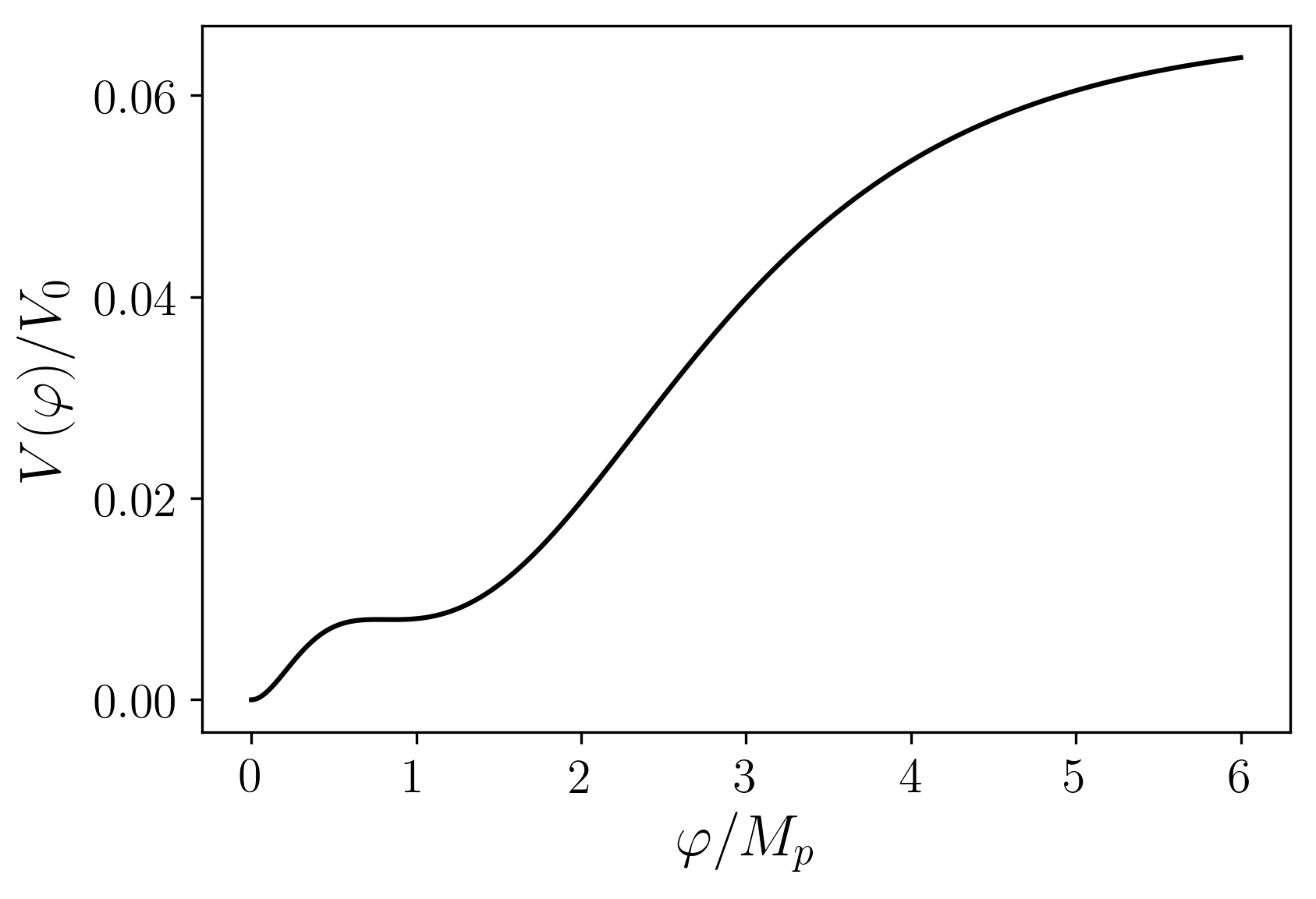}
\caption{The general shape of the potential given by Eq. \eqref{eq:potential}}
\label{fig:potential}
\end{figure}

We consider a potential constructed using inflationary $\alpha$-attractors \cite{Kallosh:2013yoa,Kallosh:2013hoa,Kallosh:2014ona} studied in \cite{Mahbub:2019uhl}. In deriving $\alpha$-attractor inflation from supergravity, one considers a supersymmetric Lagrangian with two chiral superfields $\Phi$ and $S$- the first being related to the inflaton and the second a stabilizer field along which the inflaton trajectory is stabilized. Upon stabilization, this model predicts a general class of inflaton potentials of the form
\begin{equation}
	V(\varphi)=f^{2}\left( \tanh\frac{\varphi}{\sqrt{6\alpha}} \right)
\end{equation}
where $f$ can be any holomorphic function of $\varphi\sim \text{Re}\Phi$. A potential of the following form is studied
\begin{equation}\label{eq:potential}
	V(\varphi)=V_{0}\left( 1+a_{1}-e^{-a_{2}\tanh\frac{\varphi}{\sqrt{6\alpha}}}-a_{1}e^{-a_{3}\tanh^{2}\frac{\varphi}{\sqrt{6\alpha}}} \right)^2
\end{equation}
A potential of this form has a plateau near the end of inflation where a period of ultra slow-roll (USR) inflation \cite{Dimopoulos:2017ged,Cheng:2018qof} takes place, which is where the necessary amplification of $\mathcal{P}_{\zeta}$ takes place. To numerically compute the curvature power spectrum, the background evolution of the inflaton is first obtained by solving
\begin{align}\label{eq:inflaton_evolution}
\frac{\dd^{2}\varphi}{\dd N^2}+3\frac{\dd\varphi}{\dd N}-&\frac{1}{2M_{\text{pl}}^2}\left( \frac{\dd\varphi}{\dd N} \right)^{3}\nonumber\\
&+\Bigg[ 3M_{\text{pl}}^2-\frac{1}{2}\left( \frac{\dd\varphi}{\dd N} \right)^2 \Bigg]\frac{\partial\ln V}{\partial\varphi}=0
\end{align}
where $N$ is the number of $e$-folds, related to cosmic time $t$ through $\dd N=H\dd t$. Here $H=\dot{a}/a$ is the Hubble parameter. For this paper, the parameters chosen are $\alpha=1,a_{1}=-0.3008933,a_{2}=\sqrt{2/3},a_{3}=6.1548514$. The overall multiplicative parameter $V_{0}$ is set by imposing the CMB normalization at the pivot scale $\mathcal{P}_{\zeta}(k_{\star})=2.1\times 10^{-9}$ where $k_{\star}=0.05\;\text{Mpc}^{-1}$ \cite{Aghanim:2018eyx}. The value of $V_{0}$ depends on the amount of inflation realized for any parameter set and also on how many $e$-folds into inflation the pivot scale is set. There is potential for confusion as to why the parameters appear with such long strings of digits in the decimal places. In fact, this is a rather peculiar feature in inflection point models and such fine-tuning is necessary to achieve a finite period of inflation. Discussions on this can be found in Ref. \cite{Ezquiaga:2017fvi,Hertzberg:2017dkh}. The shape of the potential is illustrated in Fig. (\ref{fig:potential}) where we see that there is an inflection point close to $\varphi=1$.\\
\indent The background evolution is obtained after numerically solving Eq. \eqref{eq:inflaton_evolution} with which the evolution of the inflaton quantum fluctuations $\delta\varphi_{\bm{k}}$ is calculated. The evolution of the Fourier modes of the inflaton fluctuations is governed by the Mukhanov-Sasaki equation, which is given by

\begin{align}\label{eq:Mukhanov_Sasaki}
\frac{\dd^{2}\delta\varphi_{\bm{k}}}{\dd N^2} + (3-\epsilon_{1})\frac{\dd\varphi_{\bm{k}}}{\dd N}+\bigg[ \left( \frac{k}{aH} \right)^{2}+(3-\epsilon_{1})\frac{\partial_{\varphi\varphi}V}{V}\nonumber\\
\qquad -2\epsilon_{1}(3-\epsilon_{1}+\epsilon_{2}) \bigg]\delta\varphi_{\bm{k}}=0
\end{align}
Equation \eqref{eq:Mukhanov_Sasaki} describes the evolution of $\delta\varphi_{\bm{k}}$ from an initially subhorizon regime ($k\gg aH$) to a superhorizon one ($k\ll aH$). The functions $\epsilon_{1}$ and $\epsilon_{2}$ are the first two Hubble flow parameters given by $\epsilon_{n}=\dd\ln\epsilon_{n-1}/\dd N$ and $\epsilon_{0}=H^{-1}$. The subhorizon limit is interesting since the initial conditions for this quantum field is set in such a region. This is because a quantum field evolving in a curved background is described by a time-dependent Hamiltonian and no unique state can be found which minimizes the Hamiltonian for all times. As a result, initial conditions for $\delta\varphi_{\bm{k}}$ are usually set in the subhorizon regime where the curvature of the Hubble patch is negligible enough for the spacetime to be approximated as Minkowski. This is called the Bunch-Davies vacuum \cite{1978RSPSA.360..117B} and is characterized by

\begin{figure*}[t]
\centering
\includegraphics[scale=0.6]{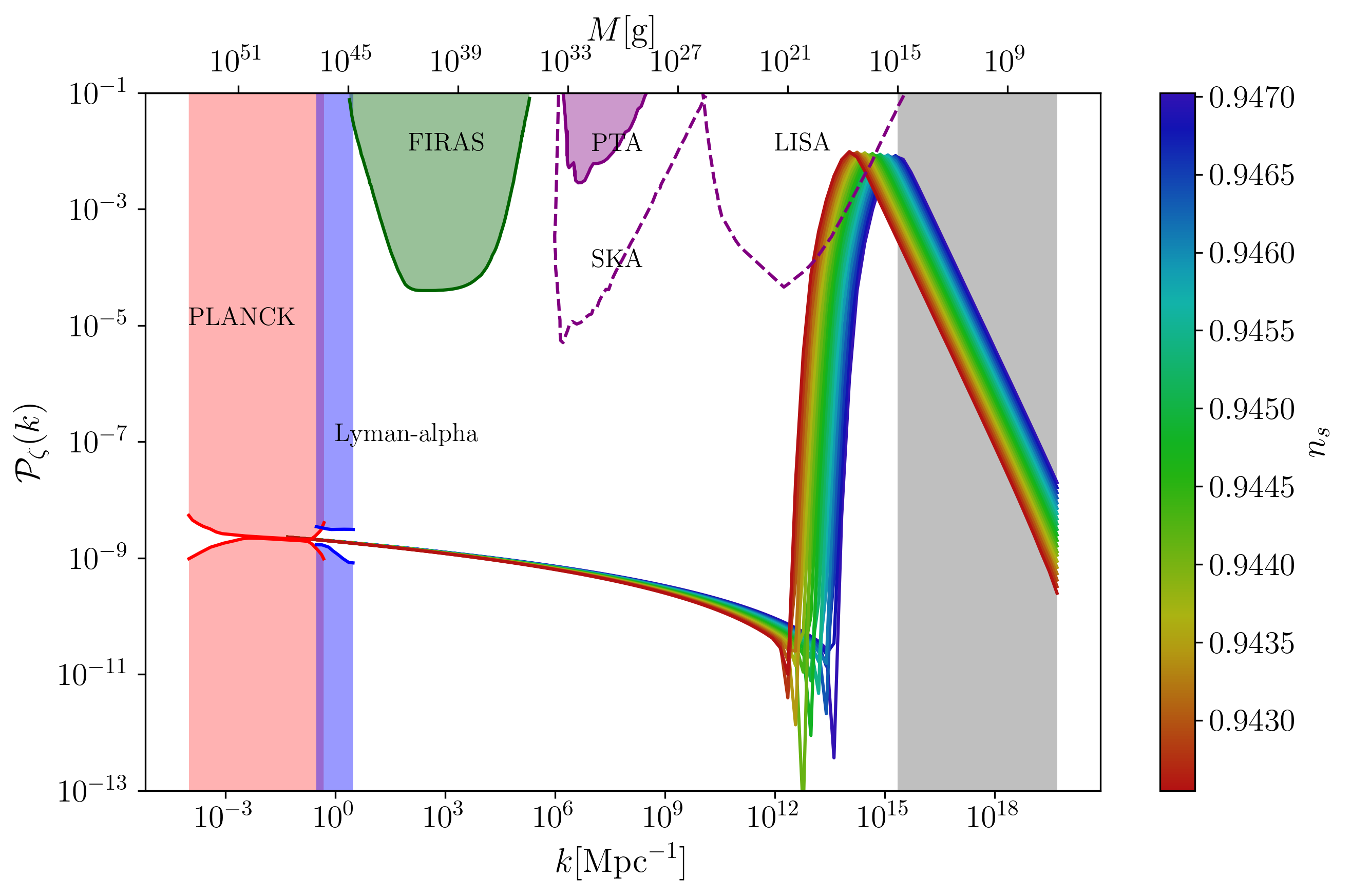}
\caption{The shape of the peaked curvature power spectrum $\mathcal{P}_{\zeta}(k)$ for different numbers of $e$-folds. The upper horizontal axis refers to the mass of a comoving Hubble patch of scale $k=aH$. The shaded regions show constraints on $\mathcal{P}_{\zeta}(k)$ arising from CMB temperature power spectrum (red) \cite{Akrami:2018odb}, Lyman-$\alpha$ forest (blue) \cite{Bird_2011}, CMB spectral distortions (green) \cite{Fixsen:1996nj} and pulsar timing array limits on gravitational waves (purple) \cite{Byrnes:2018txb}. The dashed purple lines are future constraints that can arise from gravitational wave limits set by SKA and LISA \cite{Inomata:2018epa}. The constraints on $\mathcal{P}_{\zeta}(k)$ have been obtained from \cite{bradkav,Gow:2020bzo}.}
\label{fig:power_spectrum}
\end{figure*}

\begin{align}
\delta\varphi_{k}&=\frac{1}{a\sqrt{2k}} \nonumber \\
\frac{\dd\delta\varphi_{k}}{\dd N}&=-\left( \frac{1}{a\sqrt{2k}}+i\frac{k}{aH} \frac{1}{a\sqrt{2k}}\right)
\end{align}
Once the Mukhanov-Sasaki equation is solved numerically and $\delta\varphi_{\bm{k}}$ is obtained, the power spectrum of curvature perturbations can be evaluated for the required values of $k$.
\begin{align}
\mathcal{P}_{\zeta}(k)&=\frac{k^3}{2\pi^2}\bigg\lvert\frac{\delta\varphi_{\bm{k}}}{\sqrt{2\epsilon_{1}}} \bigg\lvert^{2}_{k\ll aH}\nonumber\\
                      &=\frac{k^3}{2\pi^2}|\zeta_{\bm{k}}|^{2}_{k\ll aH}
\end{align}
where, in the spatially flat gauge, the curvature variable is related to the inflaton vacuum fluctuations by $\zeta_{\bm{k}}=\delta\varphi_{\bm{k}}/\sqrt{2\epsilon_{1}}$.\\
\indent The predictions of the model need to be compard with CMB observables at the pivot scale $k_{\star}$. Along with the CMB normalization of the power spectrum, two other quantities of interest are the scalar spectral index $n_{s}$ and the scalar to tensor ratio $r$, given by \cite{PhysRevD.98.030001}
\begin{align}
n_{s}&=1-2\epsilon_{1\star}-\epsilon_{2\star}\\
r&=16\epsilon_{1\star}
\end{align}
The `$\star$' subscripts refer to the fact that the quantities have been evaluated at the pivot scale. The curvature power spectra obtained using the potential with the chosen parameter set are shown in Fig. (\ref{fig:power_spectrum}). The colorbar refers to the value of the spectral index $n_{s}$ predicted as a result of varying the number of observable $e$-folds in the inflationary dynamics.\footnote{The number of observable $e$-folds is calculated to be the $e$-folds realized between the exit of the pivot scale $k_{\star}$ and the end of inflation.} 

The upper horizontal axis gives the mass of the comoving Hubble patch when the scale $k$ crosses the horizon. This is conventionally computed using the horizon crossing condition and is used to assign masses to PBHs which result from the collapse of perturbation modes described by $k$. The expression reads
\begin{equation}\label{eq:PBH_mass1}
M=1.6\times 10^{18}\text{g}\left( \frac{g_{\star}(T_{k})}{106.75} \right)^{-1/6}\left( \frac{k}{5.5\times 10^{13}\text{Mpc}^{-1}} \right)^{-2}
\end{equation}
where $g_{\star}$ is the number of relativistic species at the time of formation. From Fig. (\ref{fig:power_spectrum}) it is seen that, although the value of $n_{s}$ is improved when the peak is shifted towards higher values of $k$, the number of $e$-folds towards higher values of $n_{s}$ produce less massive PBHs. The grey shaded region in Fig. (\ref{fig:power_spectrum}) indicates all those PBHs which would have evaporated by now due to Hawking radiation. However, the horizon crossing condition $k=aH$ works well only when the curvature perturbations $\zeta(t,\bm{x})$ are very small and the levels which are required for PBH formation means that Eq. \eqref{eq:PBH_mass1} is not quite accurate as we shall see later. The region around the peak of the curvature power spectrum can be effectively modelled using a lognormal function of the following form
\begin{equation}\label{eq:lognormal_profile}
\mathcal{P}_{\zeta}(k)=\frac{\mathcal{A}}{\sqrt{2\pi\Delta^2}}\exp\left[ -\frac{\ln^{2}\left( k/k_{0} \right)}{2\Delta^2} \right]
\end{equation}
where $\mathcal{A}$ and $\Delta$ control the amplitude and width of the peak while $k_{0}$ denotes the location of the peak value of $\mathcal{P}_{\zeta}(k)$. The power spectra shown in Fig. (\ref{fig:power_spectrum}) can be fitted using the lognormal profile for $\mathcal{A}\approx 0.018$ and $\Delta\approx 0.8$ wiith $k_{0}$ being a free parameter. This functional form of the power spectrum will be used in the later chapters when peaks theory is used to compute PBH abundance.

\section{PBH formation criterion}
PBHs form from the collapse of overdense regions. In the simplest terms, this collapse criterion is said to be met when these overdense regions exceed some predetermined threshold $\delta_{\text{th}}$. One of the earliest, and certainly one of the simplest, estimate of this was obtained by Bernard Carr using simple Jeans stability analysis, where this threshold was calculated to be $\delta_{\text{th}}\approx 1/3$ \cite{Carr:1975qj}. Other, more refined, values of the threshold were obtained from numerical simulations of matter collapsing in the early Universe \cite{Nakama:2013ica,Niemeyer:1999ak,Escriva:2019phb,Harada:2013epa,Yoo:2020lmg}. \\
\indent In the study of PBHs from the collapse of overdense regions, the relationship between the primordial curvature perturbations $\zeta(t,\bm{x})$ and the overdensity of matter in the post-inflationary universe $\delta(t,\bm{x})$ is used, which is highly nonlinear \cite{Salopek:1990jq,PhysRevD.60.084002,Harada:2015yda,Germani:2018jgr}

\begin{equation}\label{eq:density}
	\delta(t,\bm{x})=-\frac{4(1+w)}{3w+5}\frac{1}{(aH)^2}e^{5\zeta(t,\bm{x})/2}\grad^{2}e^{-\zeta(t,\bm{x})/2}
\end{equation}
The quantity $w$ is the equation of state parameter and is equal to $1/3$ for radiation. The nonlinear relationship between $\delta$ and $\zeta$ itself has consequences for PBH formation, especially in relation to abundance, which has been studied in \cite{DeLuca:2019qsy,Kawasaki:2019mbl}. An accurate determinant of the threshold for collapse is set using the compaction function. This is defined as
\begin{equation}
	\mathcal{C}=\frac{\delta M}{R}
\end{equation}
where $\delta M$ is the excess Misner-Sharp mass enclosed within a sphere of areal radius $R$. The Misner-Sharp mass is given by the expression \cite{PhysRev.136.B571}
\begin{equation}
M_{\text{MS}}(r)=4\pi\int_{0}^{r}\dd r'\;\rho(r')R^{2}(r')\dv{R(r')}{r'}
\end{equation}
The areal radius depends on the particular way the FLRW metric is paramterized and spherical symmetry allows the Universe to be described by the following metric
\begin{equation}
\dd s^2 = -\dd t^2 + a^{2}(t)e^{-2\zeta(r)}\left( \dd r^2 + r^2 \dd\Omega^2 \right)
\end{equation}
The $\delta M$ then represents how much excess $M_{\text{MS}}$ is contained within a spherical region of areal radius $R(r,t)=a(t)re^{-\zeta(r)}$ compared to the background mass given by $\bar{M}=4\pi R^{3}\bar{\rho}/3$. Calculating the mass excess
\begin{align}
\delta M&=4\pi\int_{0}^{r}\dd r'\;\bigg[ \rho(r')a^{2}r'^{2}e^{-2\zeta(r')}\frac{\dd}{\dd r'}\left(ar'e^{-\zeta(r')}\right) \nonumber\\
& \qquad \qquad\qquad\qquad\qquad -\bar{\rho}a^{3}r'^2\bigg]\nonumber\\
&=\frac{3}{2}H^{2}\int_{0}^{r}\dd r'\;r'^{2}(1+\delta)a^{3}e^{-3\zeta(r')}\left( 1-r'\frac{\dd \zeta(r')}{\dd r'} \right)\nonumber\\
& \qquad \qquad\qquad\qquad\qquad -\frac{1}{2}H^{2}a^{3}r^{3}
\end{align}
We note that the areal radius of the homogeneous FLRW universe is $R=ar$ and the curvature term does not appear. In the long wavelength limit, where Eq. \eqref{eq:density} is valid, the compaction function takes a relatively simple form
\begin{equation}
\mathcal{C}=\frac{1}{2}\bar{\delta}(HR)^2 
\end{equation}
where $\bar{\delta}$ is the volume-averaged overdensity within the radius $r$. In the comoving gauge, the compaction function takes a simple form
\begin{equation}
\mathcal{C}(r)=\frac{1}{3}\left[ 1-\left( 1-r\frac{\dd\zeta}{\dd r} \right)^2 \right]
\end{equation}
Using this definition, the lengthscale relevant for collapse, $r_{m}$, is computed by maximizing the compaction function and collapse is said to take place if $\mathcal{C}_{\text{max}}>\mathcal{C}_{\text{th}}$. As a consequence, we see that the requirement for gravitational collapse is not simply a threshold density but dependent on the shape of the perturbations in real space.
\section{PBH abundance from optimized peaks theory}
This chapter discusses the calculation of PBH abundance using peaks theory. Detailed derivations can be found in \cite{Yoo:2018kvb,1986ApJ...304...15B,2010gfe..book.....M}.
\subsection{Thresholds and PBH mass}
In this work, we assume that the curvature perturbations are Gaussian random fields which are exactly specified through a two-point correlation function
\begin{equation}
\langle \zeta(\bm{k})\zeta^{*}(\bm{k}') \rangle=\frac{2\pi^2}{k^3}\mathcal{P}_{\zeta}(k)(2\pi)^{3}\delta^{(3)}\left( \bm{k}-\bm{k}' \right)
\end{equation}
where $\mathcal{P}_{\zeta}$ is the power spectrum that has been defined previously. One can also define the moments of the power spectrum using
\begin{equation}\label{eq:Pzeta_moments}
\sigma_{n}^{2}=\int_{\mathbb{R}^{+}}\dd\ln k\;k^{2n}\mathcal{P}_{\zeta}(k)
\end{equation}
For the lognormal power spectrum, using the variable $u=k/k_{0}$, these moments are given by the following
\begin{align}\label{eq:power_spectrum_moments}
\sigma_{n}^{2}&=\frac{\mathcal{A}}{\sqrt{2\pi\Delta^{2}}}k_{0}^{2n}\int_{0}^{\infty}\dd u\;u^{2n-1}e^{-\ln^{2}u/2\Delta^2}\nonumber\\
&=\mathcal{F}(n,\Delta)\mathcal{A}k_{0}^{2n}
\end{align}
where
\begin{align}\label{eq:spectral}
\mathcal{F}(n,\Delta)&=\frac{1}{\sqrt{2\pi\Delta^{2}}}\int_{0}^{\infty}\dd u\;u^{2n-1}e^{-\ln^{2}u/2\Delta^2}\nonumber\\
                     &=\frac{1}{\sqrt{2\pi\Delta^{2}}}\int_{-\infty}^{\infty}\dd y\;e^{-y^{2}/2\Delta^{2}+2ny}\nonumber\\
                     &=e^{2n^{2}\Delta^{2}}
\end{align}

\begin{table}[]
\begin{tabular}{@{}llll@{}}
\toprule
                            & $\Delta=0.6$                 & $\Delta=0.7$                 & $\Delta=0.8$                 \\ \midrule
\multicolumn{1}{|l|}{$n=1$} & \multicolumn{1}{l|}{2.05443} & \multicolumn{1}{l|}{2.66446} & \multicolumn{1}{l|}{3.59664} \\ \midrule
\multicolumn{1}{|l|}{$n=2$} & \multicolumn{1}{l|}{17.8143} & \multicolumn{1}{l|}{50.4004} & \multicolumn{1}{l|}{167.335} \\ \midrule
\multicolumn{1}{|l|}{$n=3$} & \multicolumn{1}{l|}{651.971} & \multicolumn{1}{l|}{6768.26} & \multicolumn{1}{l|}{100710}  \\ \bottomrule
\end{tabular}
\caption{The numerical values of $\mathcal{F}(n,\Delta)$ for the first three moments of $\mathcal{P}_{\zeta}$ for $\Delta=0.6,0.7$ and $0.8$.}
\label{table:table1}
\end{table}

\begin{figure*}[t]
\centering
\includegraphics{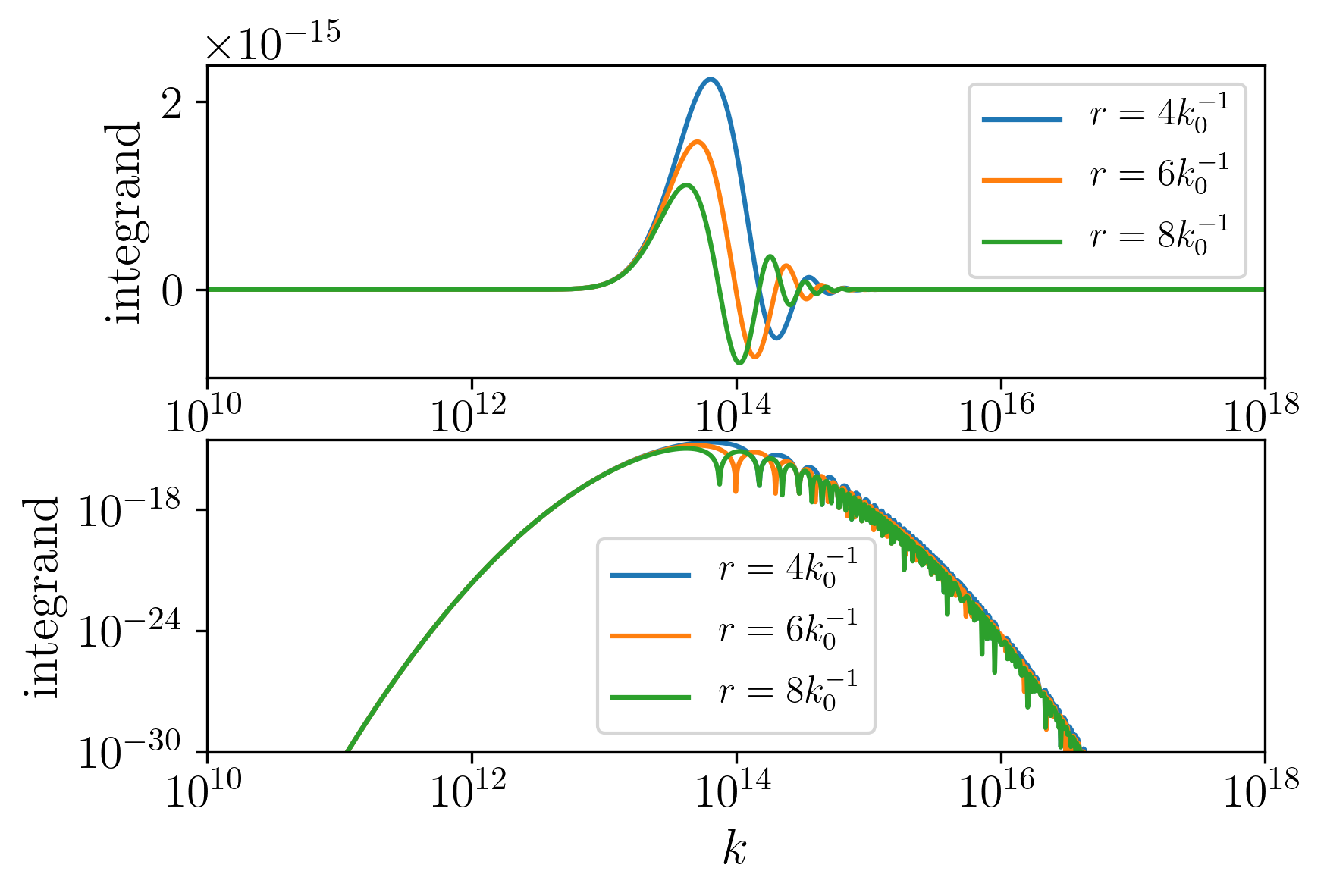}
\caption{The integrand of Eq. \eqref{eq:psi}, showing oscillatory behavior around the peak centered at $k_{0}$. This shows that the limits of integration turn out to be confined to a small interval.}
\label{fig:integrand}
\end{figure*}

\begin{figure}[h]
\centering
\includegraphics[scale=0.6]{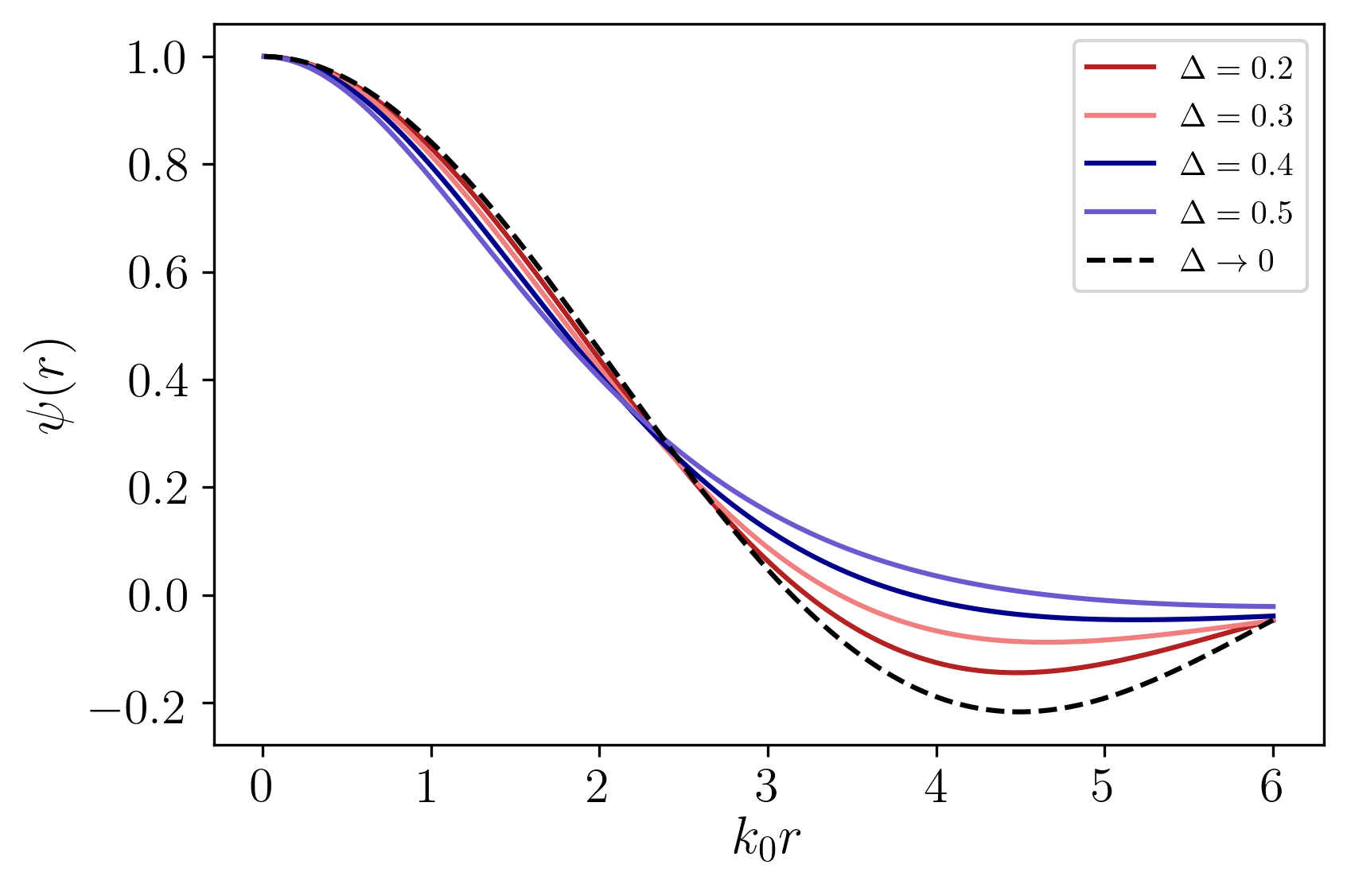}
\caption{The typical shape of $\psi(r)$ obtained using the lognormal power spectrum for different values of $\Delta$. The dashed, black line represents a Dirac delta function-type $\mathcal{P}_{\zeta}(k)$, which is expected to be recovered when $\Delta\rightarrow 0$.}
\label{fig:psi_profiles}
\end{figure}

In the second line of Eq. \eqref{eq:spectral}, the change of variable $\ln u=y$ was used to convert it to a Gaussian integral. Numerical values of $\mathcal{F}(n,\Delta)$ relevant to this work are given in Table (\ref{table:table1}). Equation \eqref{eq:power_spectrum_moments} can be numerically solved to obtain the moments of the power spectrum for any well-behaved profile describing $\mathcal{P}_{\zeta}(k)$, however, analytical expressions might not exist for a large class of them. In peaks theory, structure is formed from peaks in the curvature and overdensity fields. Typically, in the high peak limit, the average profile of the curvature perturbations can be described by two variables $k_{*}$ and $\mu$ with $\mu$ controlling the amplitude and $1/k_{*}$ controlling the curvature scale of $\zeta$. Then, average curvature profile can be written as
\begin{equation}
\frac{\bar{\zeta}(r,k_{*})}{\mu}\equiv g(r;k_{*})=g_{0}(r)+k_{*}^{2}g_{1}(r)
\end{equation}
where
\begin{align}
g_{1}(r)&=-\frac{1}{1-\gamma^2}\left( \psi(r)+\frac{1}{3}R_{*}^2\grad^{2}\psi(r) \right)\label{eq:g1}\\
g_{2}(r)&=\frac{1}{\gamma(1-\gamma^2)}\frac{\sigma_{0}}{\sigma_{2}}\left( \gamma^{2}\psi(r)+\frac{1}{3}R_{*}^2\grad^{2}\psi(r) \right)\label{eq:g2}
\end{align}

Here, $\gamma=\sigma_{1}^{2}/(\sigma_{0}\sigma_{2})$ is closely related to the sharpness of the profile. For example, in the case where $\Delta\rightarrow 0$ in Eq. \eqref{eq:lognormal_profile} (representing a Dirac delta function), the parameter $\gamma\rightarrow\infty$. The other parameter is defined as $R_{*}=\sqrt{3}\sigma_{1}/\sigma_{2}$ (which is related to the curvature scale, $k_{*}^{-1}$). Furthermore, Eq. \eqref{eq:g1} and \eqref{eq:g2} depend on $\psi(r)$ which is defined as
\begin{equation}\label{eq:psi}
\psi(r)=\frac{1}{\sigma_{0}^2}\int_{\mathbb{R}^{+}}\dd \ln k\frac{\sin kr}{kr}\mathcal{P}_{\zeta}(k)
\end{equation}

\begin{figure*}[t]
\centering
\includegraphics[scale=0.6]{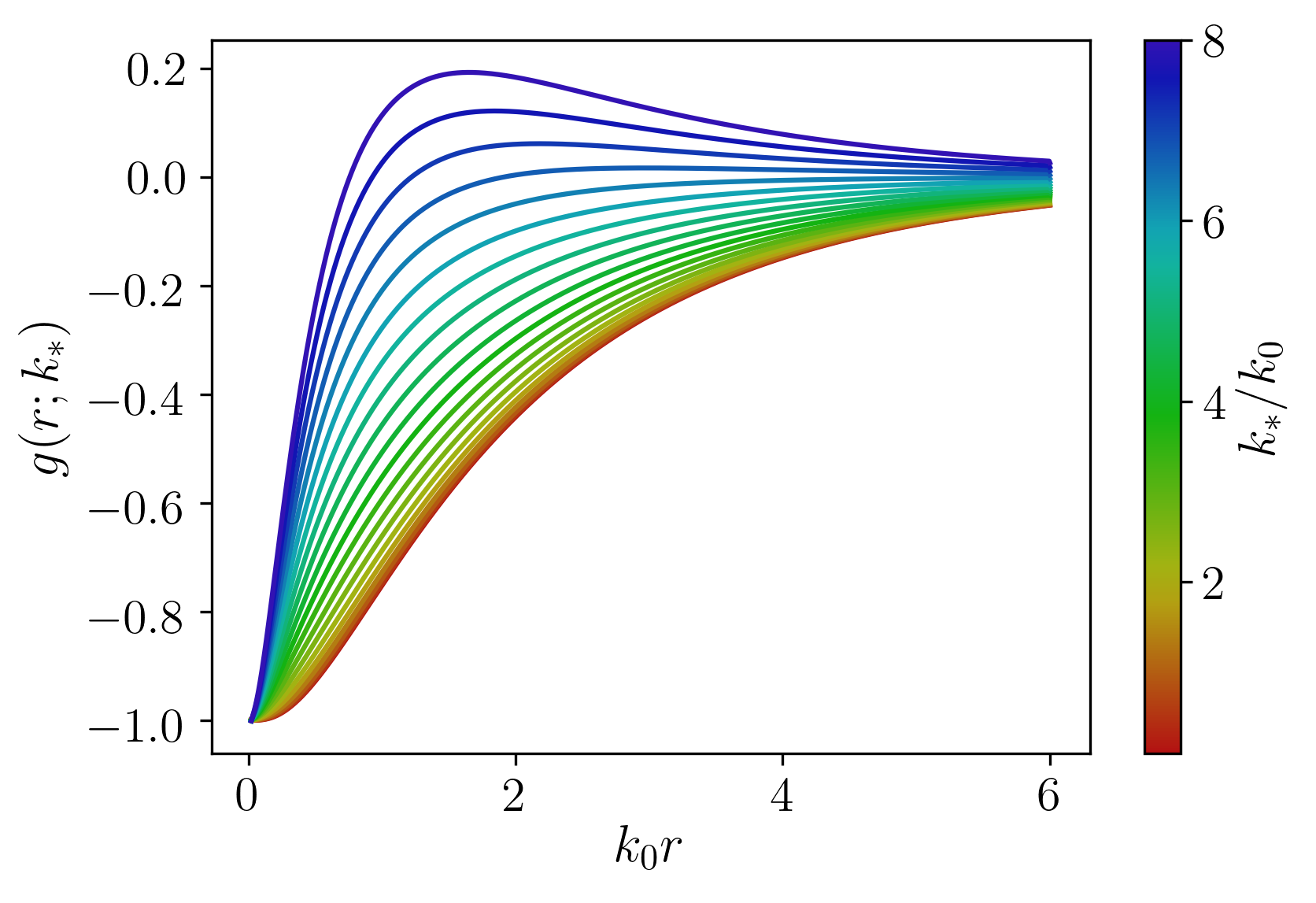}
\includegraphics[scale=0.6]{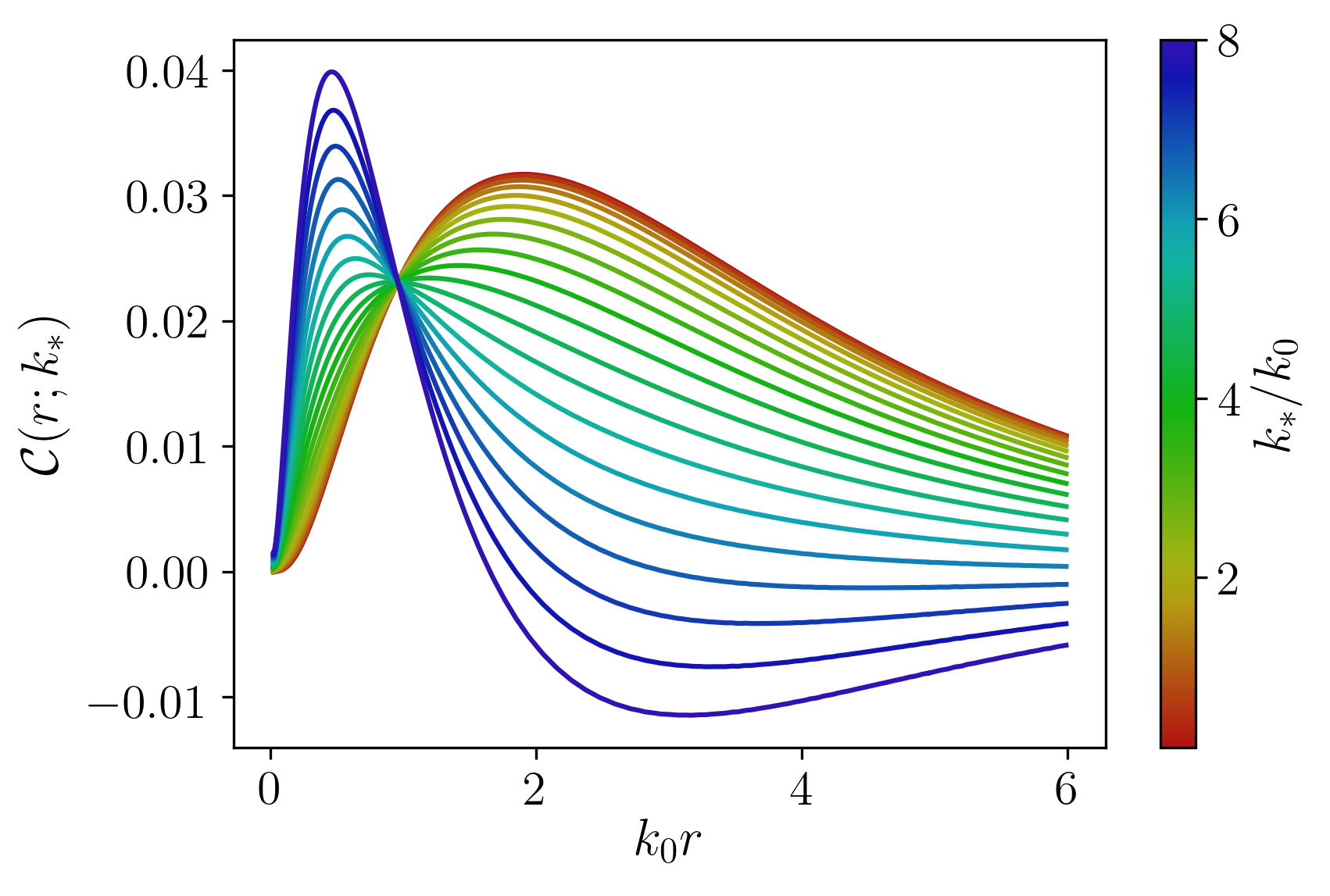}
\caption{Plots of the average curvature profile $g(r;k_{*})$ (left panel) and compaction function $\mathcal{C}(r;k_{*})$ (right panel) for different values of curvature scale in relation to the peak amplitude wavenumber $k_{*}/k_{0}$. From the plot on the right, we can identify where $\mathcal{C}_{\text{max}}$ occurs for a given $k_{*}$ and compute the lengthscale necessary for collapse $r_{m}$.}
\label{fig:compaction}
\end{figure*}

The integrand in Eq. \eqref{eq:psi} is quite oscillatory in a small interval centered around the peak, $k_{0}$, which rapidly decays on both sides. This can be best seen in Fig. (\ref{fig:integrand}). The rapid decay on either side of the oscillatory band means that the limits of integration becomes rather small and the upper limit, $k_{\text{max}}$, need not be arbitrarily large. As a result, the integral can be evaluated as
\begin{equation}
\psi(r)\simeq \frac{1}{\sigma_{0}^2}\int_{\epsilon}^{k_{\text{max}}\sim10^{a}k_{0}}\dd \ln k\frac{\sin kr}{kr}\mathcal{P}_{\zeta}(k)
\end{equation}
where $10^{a}k_{0}$ on the upper limit of the integral refers to some value a few orders of magnitude larger than $k_{0}$ and $\epsilon$ is some small lower limit. For this work, the upper limit has been chosen to be $k_{\text{max}}=10^{30}\sim 10^{16}k_{0}$. Figure (\ref{fig:psi_profiles}) illustrates the typical shape of $\psi(r)$ that is obtained from the lognormal profile of $\mathcal{P}_{\zeta}(k)$ and has been plotted for different values of $\Delta$. In the limit where $\mathcal{P}_{\zeta}(k)\sim \delta(k-k_{0})$, $\psi(r)\sim \frac{\sin k_{0}r}{k_{0}r}$, shown in dashed black line. In \cite{Yoo:2018kvb}, Eq. \eqref{eq:Pzeta_moments}-\eqref{eq:psi} could be analytically solved due to the particular functional form of $\mathcal{P}_{\zeta}(k)$ chosen. However, it did not allow for much flexibility in terms of controlling the height and width of the peak which can be easily accomplished using a lognormal function. The drawback, however, is that these equations cannot be solved analytically. In the numerical evaluation of these equations, the amplitude of the lognormal power spectrum $\mathcal{A}$ is identified with the zeroth moment of $\mathcal{P}_{\zeta}(k)$ such that $\mathcal{A}=\sigma_{0}^2$. Although the functions $g(r;k_{*})$ and $\psi(r)$ are functions of $r$, their dependence on the lengthscale is made dimensionless by considering the combination $k_{0}r$, even though there is no a priori reason why one should consider it. Using these considerations, the average curvature profile and compaction function have been plotted in Fig. (\ref{fig:compaction}), showing their dependence on $k_{*}$.\\
\begin{figure*}[t]
\centering
\includegraphics[scale=0.6]{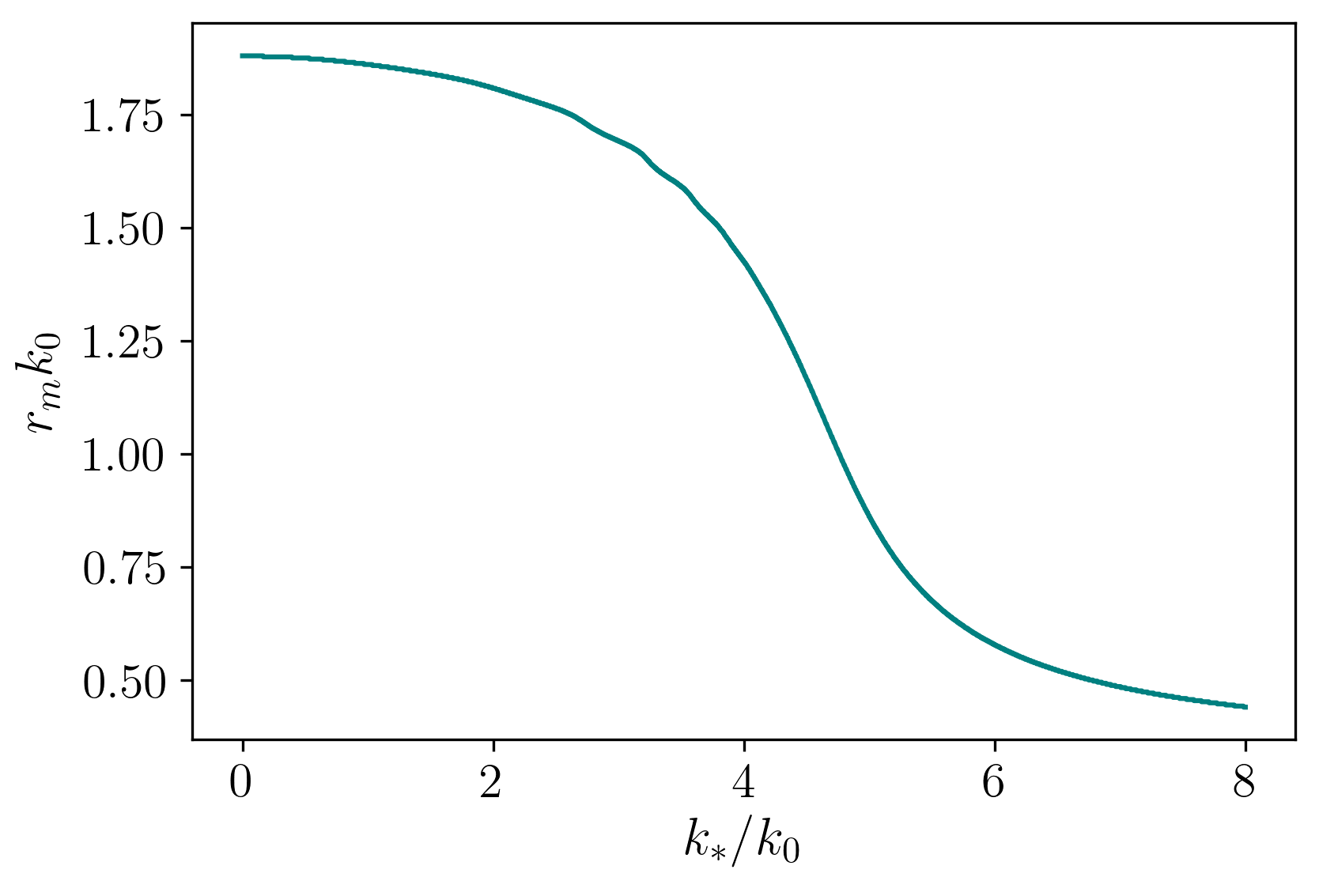}
\includegraphics[scale=0.6]{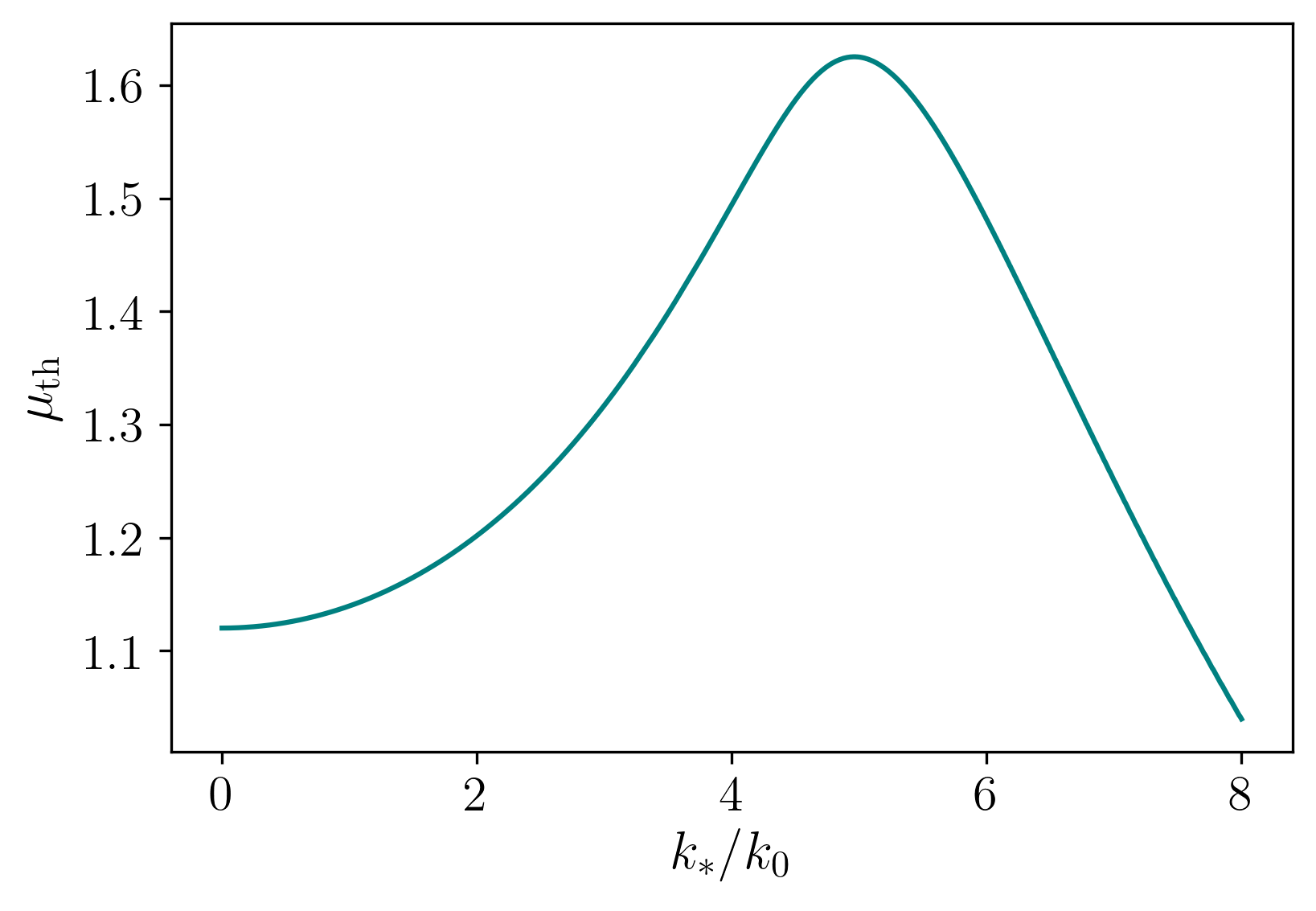}
\caption{Plots of the radius of maximization of $\mathcal{C}$ (left panel) and $\mu_{\text{th}}$ (right panel) as a function of $k_{*}$. These are then converted into functions of $M$ in the final calculations.}
\label{fig:rm}
\end{figure*}
\indent Using the definition of $\mathcal{C}(r)$, the threshold value of $\mu$ is obtained
\begin{equation}
\mu_{\text{th}}(k_{*})=\frac{1-\sqrt{1-3\mathcal{C}_{\text{th}}}}{r_{m}(k_{*})g'_{m}(r_{m};k_{*})}
\end{equation}
In Fig. (\ref{fig:rm}), the radius at which the compaction function is maximized, $r_{m}$ (left panel), and $\mu_{\text{th}}$ (right panel) are shown as a function of $k_{*}$. Since PBH abundance is generally expressed as a function of mass, one needs to consider how $\mu$ is related to $M$. Here, a more accurate computation of PBH mass can be formulated and compared with the one given by Eq. \eqref{eq:PBH_mass1}. As usual, the starting assumption is that PBHs are assigned a mass that is a fraction of the collapsed Hubble patch upon re entry of high curvature modes. The usual horizon crossing definition of $k=aH$ is really an approximation in the limit where $\zeta(r)$ is small. The horizon crossing condition, in reality, is expressed in real space and is given by $H(t,r)R(t,r)=1$. At the scale $r_{m}$ where the compaction function is maximized, this turns out to be
\begin{equation}\label{eq:horizon_cross}
aH=\frac{1}{r_{m}}e^{\mu g_{m}}
\end{equation}
It can be seen that, when $g_{m}$ is small (such that curvature is small), $r_{m}^{-1}\sim k$ and the usual horizon crossing condition is recovered. In the case of PBH formation, where large curvature perturbation becomes important, the factor $e^{\mu g_{m}}$ will produce a nontrivial contribution towards the mass of the PBH. With the assumption that a fraction $\gamma$ of the horizon mass goes into PBHs, $M=\gamma 4\pi\rho R^{3}/3$. Using the definition of $H$ and invoking the horizon crossing condition, the mass can be written as $M=\gamma/2H$. Then
\begin{equation}\label{eq:PBH_mass2}
M(\mu,k_{*})=\frac{1}{2}\gamma ar_{m}e^{-\mu g_{m}}=\gamma M_{\text{eq}}k_{\text{eq}}^{2}r_{m}^{2}e^{-2\mu g_{m}}
\end{equation}
where $M_{\text{eq}}$ and $k_{\text{eq}}$ are the mass of the Hubble horizon and comoving wavenumber at matter-radiation equality, which are both known quantities. Here we see that the PBH depends on $\zeta(r)$ through the term $e^{-2\mu g_{m}}$ and the lengthscale at which the compaction function is maximized, $r_{m}$. For an extended power spectrum, the value of $\mu$ maybe bounded from below for constant $M$. As a result, the region for PBH formation is given by
\begin{equation}
\mu > \mu_{\text{b}}=\text{max}\{\ \mu_{\text{min}}(M),\mu_{\text{th}}(M) \}\
\end{equation}
\subsection{PBH abundance and CDM fraction}\label{sec:PBH_abundance}
Since a key assumption states that the curvature perturbations are Gaussian, their distrubution is given by a multidimensional Gaussian probability density functional (PDF) of the form
\begin{align}
P(\zeta(\bm{x}_{1}),...,\zeta(\bm{x}_{n}))\dd \bm{\zeta}&=\frac{1}{\left( 2\pi \right)^{n/2}|\text{det}\bm{\Sigma}|^{1/2}}\nonumber\\
&\times\exp\left[ -\frac{1}{2}\bm{\zeta}_{\text{I}}\left( \bm{\Sigma}^{-1} \right)^{\text{IJ}}\bm{\zeta}_{\text{J}} \right]\dd \bm{\zeta}
\end{align}
where the indices $\text{I}$ and $\text{J}$ refer to the $n$ different positions where $\zeta$ is evaluated in matrix form, eg. $\bm{\zeta}^{\text{T}}=\left( \zeta(x_{1}),...,\zeta(x_{n}) \right)$. The components of the matrix $\bm{\Sigma}$ are $\bm{\Sigma}_{\text{IJ}}=\langle \zeta(\bm{x}_{\text{I}})\zeta^{*}(\bm{x}_{\text{J}}) \rangle$ and, hence, related to the power spectrum. Considering the curvature profile $\zeta(x_{i})$ and expanding it to second order and centering the peak at the origin, 
\begin{align}
\zeta(x_{k})&=\zeta(0)+\left(\partial^{i}\zeta\right) x_{i}+\frac{1}{2}\left(\partial^{i}\partial^{j}\zeta\right) x_{i}x_{j}+\mathcal{O}\left( x^3 \right)\nonumber\\
&\approx \zeta_{0}+\zeta^{i}_{1}x_{i}+\frac{1}{2}\zeta^{ij}_{2}x_{i}x_{j}
\end{align}
The  $\zeta^{ij}_{2}$ can be packed as a symmetric $3\times 3$ matrix with six independent components. The only nonvanishing correlations between $\zeta_{0}$, $\zeta^{i}_{1}$ and $\zeta^{ij}_{2}$ are
\begin{align}
\langle \zeta_{0}\zeta_{0} \rangle&=\sigma_{0}^{2}\\
\langle \zeta_{0}\zeta^{ij}_{2} \rangle&=-\frac{\sigma_{1}^{2}}{3}\delta^{ij}=-\langle \zeta^{i}_{1}\zeta^{j}_{1} \rangle\\
\langle \zeta^{ij}_{2}\zeta^{kl}_{2} \rangle&=\frac{\sigma_{2}^{2}}{15}\left( \delta^{ij}\delta^{kl}+\delta^{ik}\delta^{jl}+\delta^{il}\delta^{kj} \right)
\end{align}
The multidimensional PDF can be conveniently expressed using the following variable redefinitions
\begin{align}
\nu&=-\frac{\zeta_{0}}{\sigma_{0}}\\
\eta^{i}&=\frac{\zeta^{i}_{1}}{\sigma_{1}}\\
\xi_{1}&=\frac{1}{\sigma_{2}}\left( \lambda_{1}+\lambda_{2}+\lambda_{3} \right)\\
\xi_{2}&=\frac{1}{2\sigma_{2}}\left( \lambda_{1}-\lambda_{2} \right)\\
\xi_{3}&=\frac{1}{2\sigma_{2}}\left( \lambda_{1}-2\lambda_{2}+\lambda_{3} \right)
\end{align}
where the $\lambda_{i}$'s occuring in the preceeding equations are the eigenvalues of the $-\zeta^{ij}_{2}$ matrix with $\lambda_{1}\geq\lambda_{2}\geq\lambda_{3}$. These now represent the new degrees of the freedom of the PDF which can be written as $P(\nu,\bm{\xi},\bm{\eta})\dd\nu\dd\bm{\xi}\dd\bm{\eta}=P_{1}(\eta,\xi_{1})P_{2}(\xi_{2},\xi_{3},\bm{\eta})\dd\nu\dd\bm{\xi}\dd\bm{\eta}$ with
\begin{align}\label{eq:prob1}
P_{1}(\nu,\xi_{1})\dd\nu\dd\xi_{1}&=\frac{\mu}{2\pi\sigma_{0}\sigma_{2}}\frac{1}{\sqrt{1-\gamma^2}}\nonumber\\
&\times\exp\bigg[ -\frac{1}{2}\mu^2\bigg( \frac{1}{\sigma_{0}^2} +\frac{1}{\sigma_{2}^2}\frac{\left( k_{*}^2 -k_{c}^2 \right)}{1-\gamma^2} \bigg) \bigg]\dd\mu\dd k_{*}^2 \\
&=\frac{\mu}{2\pi\sigma_{0}\sigma_{2}}\frac{1}{\sqrt{1-\gamma^2}}e^{-\mu^{2}/2\tilde{\sigma}^{2}(k_{*})}\dd\mu\dd k_{*}^2 
\end{align}
and,
\begin{align}
P_{2}(\bm{\eta},\xi_{2},\xi_{3})&\dd\xi_{2}\dd\xi_{3}\dd\bm{\eta}=\frac{5^{5/2}3^{7/2}}{(2\pi)^2}\xi_{2}\left( \xi_{2}^2-\xi_{3}^2 \right)\nonumber\\
&\times\exp\left[ -\frac{5}{2}\left( 3\xi_{2}^2 +\xi_{3}^2 \right) \right]\nonumber\\
&\times \exp\left[ -\frac{3}{2}\eta_{i}\eta^{i} \right]\dd\xi_{2}\dd\xi_{3}\dd\bm{\eta}
\end{align}
In Eq. \eqref{eq:prob1}, $\mu=\nu\sigma_{0}$, $k_{*}^2=\xi_{1}\sigma_{2}/\mu$ and $k_{c}=\sigma_{1}/\sigma_{0}$. Using the expressions for $P_{1}$ and $P_{2}$, peaks theory can be applied to compute the number of extrema in $\zeta$ that occur in a comoving volume. The number density of $\zeta$ extrema in the space of $(\bm{x}, \nu,\xi_{1})$ is defined as
\begin{align}\label{eq:next}
&n_{\text{ext}}(\bm{x},\nu,\xi_{1})\dd\bm{x}\dd\nu\dd\xi_{1}\nonumber\\
&=\sum_{p}\delta^{(3)}\left( \bm{x}-\bm{x}_{p} \right)\delta\left( \nu-\nu_{p} \right)\delta\left( \xi_{1}-\xi_{1p} \right)\dd\bm{x}\dd\nu\dd\xi_{1}
\end{align}
The extremal values of each variable have been labelled with a subscript `$p$' and $\bm{x}_{p}$ is used to denote the position where the extrema occur. Since these are extrema of the curvature perturbations, $\partial^{i}\zeta=\zeta^{i}_{1}=\eta=0$. As a result, around each extremum, the curvature perturbation can be expressed as $\zeta(\bm{x})\sim \zeta(\bm{x}_{p})+\frac{1}{2}\zeta^{ij}_{2}\left( \bm{x}-\bm{x}_{p} \right)_{i}\left( \bm{x}-\bm{x}_{p} \right)_{j}$. With this, one can show that $\eta^{i}\equiv \partial^{i}\zeta=\partial^{i}\partial^{j}\zeta|_{\bm{x}=\bm{x}_{p}}(\bm{x}-\bm{x}_{p})_{j}$. Provided that $\partial^{i}\partial^{j}\zeta|_{\bm{x}=\bm{x}_{p}}$ is a nonsingular matrix
\begin{align}
\delta^{(3)}\left( \bm{x}-\bm{x}_{p} \right)&=\text{det}|\partial^{i}\partial^{j}\zeta|_{\bm{x}=\bm{x}_{p}}\delta^{(3)}\left( \bm{\eta} \right)\nonumber\\
&=\frac{|\lambda_{1}\lambda_{2}\lambda_{3}|}{\sigma_{1}^{3}}\delta^{(3)}\left( \bm{\eta} \right)
\end{align}
where, at $\bm{x}=\bm{x}_{p}$, $\bm{\eta}=0$ and $27\lambda_{1}\lambda_{2}\lambda_{3}=\left[ \left( \xi_{1}+\xi_{3} \right)^2 -9\xi_{2}^2\right]\left( \xi_{1}-2\xi_{3} \right)\sigma_{2}^3$. Next, the number density of structure forming peaks should be consider, which is a bit more restrictive than $n_{\text{ext}}$ given by Eg. \eqref{eq:next}. As a result, the additional condition of the positivity of the $\lambda_{3}$ eigenvalue should be imposed since for a peak $\zeta^{ij}_{2}$ has to be negative definite. Hence, the peak number density $n_{\text{pk}}$ is constructed as follows
\begin{align}\label{eq:npk1}
n_{\text{pk}}(\nu,\xi_{1})\dd\nu\dd\xi_{1}&=\langle n_{\text{ext}}\Theta(\lambda_{3}) \rangle\dd\nu\dd\xi_{1}\nonumber\\
&=\left( \frac{3}{2\pi} \right)^{3/2}\left( \frac{\sigma_{2}}{\sigma_{1}} \right)^{3}f(\xi_{1})P_{1}(\nu,\xi_{1})\dd\nu\dd\xi_{1}
\end{align}
where,
\begin{align}
f(\xi_{1})&=\frac{1}{2}\xi_{1}(\xi_{1}^{2}-3)\left[ \erf\left( \frac{1}{2}\sqrt{\frac{5}{2}}\xi_{1} \right)+\erf\left( \sqrt{\frac{5}{2}}\xi_{1} \right) \right]\nonumber\\
&\sqrt{\frac{2}{5\pi}}\left[ \left( \frac{8}{5}+\frac{31}{4}\xi_{1}^2 \right)e^{-5\xi^{2}/8}+\left( -\frac{8}{5}+\frac{1}{2}\xi_{1}^2 \right)e^{-5\xi^{2}/2} \right]
\end{align}
The number of peaks in Eq. \eqref{eq:npk1} must be expressed in terms of the PBH mass. Using the fact that $\nu=\mu/\sigma_{0}$ and $\xi_{1}=\mu k_{*}^{2}/\sigma_{2}$ and using the fact that, from Eq. \eqref{eq:PBH_mass2}
\begin{equation}
\mu(M,k_{*})=-\frac{1}{2g_{m}}\ln\left( \frac{1}{k_{\text{eq}}^{2}r_{m}^{2}}\frac{M}{M_{\text{eq}}} \right)
\end{equation}
the number density of peaks can be expressed in terms of $\mu$ and $M$
\begin{align}
n_{\text{pk}}\left( \mu,M \right)&\dd\mu\dd M =\left( \frac{3}{2\pi} \right)^{3/2}\frac{\sigma_{2}^2}{\sigma_{0}\sigma_{1}^{3}}\mu k_{*}f\left( \frac{\mu k_{*}^2}{\sigma_{2}} \right)\nonumber\\
&\times P_{1}\left( \frac{\mu}{\sigma_{0}},\frac{\mu k_{*}^2}{\sigma} \right)\bigg\lvert\frac{\dd \ln r_{m}}{\dd k_{*}}-\mu\frac{\dd g_{m}}{\dd k_{*}} \bigg\lvert^{-1}\dd\mu\dd\ln M
\end{align}
Using this, the PBH formation fraction can be approximated as
\begin{align}\label{eq:beta1}
\beta(M)&\approx 2\gamma\left( \frac{3}{2\pi} \right)^{1/2}k_{\text{eq}}^{-3}\frac{\sigma_{2}^2}{\sigma_{0}\sigma_{1}^3}\left( \frac{M}{M_{\text{eq}}} \right)^{3/2}\nonumber\\
&\times\bigg[ \tilde{\sigma}^{2}(k_{*})k_{*}f\left( \frac{\mu k_{*}^2}{\sigma_{2}} \right)P_{1}\left( \frac{\mu}{\sigma_{0}},\frac{\mu k_{*}^2}{\sigma_{2}} \right)\nonumber\\
&\times\ \bigg\lvert\frac{\dd \ln r_{m}}{\dd k_{*}}-\mu\frac{\dd g_{m}}{\dd k_{*}} \bigg\lvert^{-1} \bigg]_{\mu=\mu_{\text{b}}}
\end{align}
\begin{figure*}[t]
\centering
\includegraphics[scale=0.55]{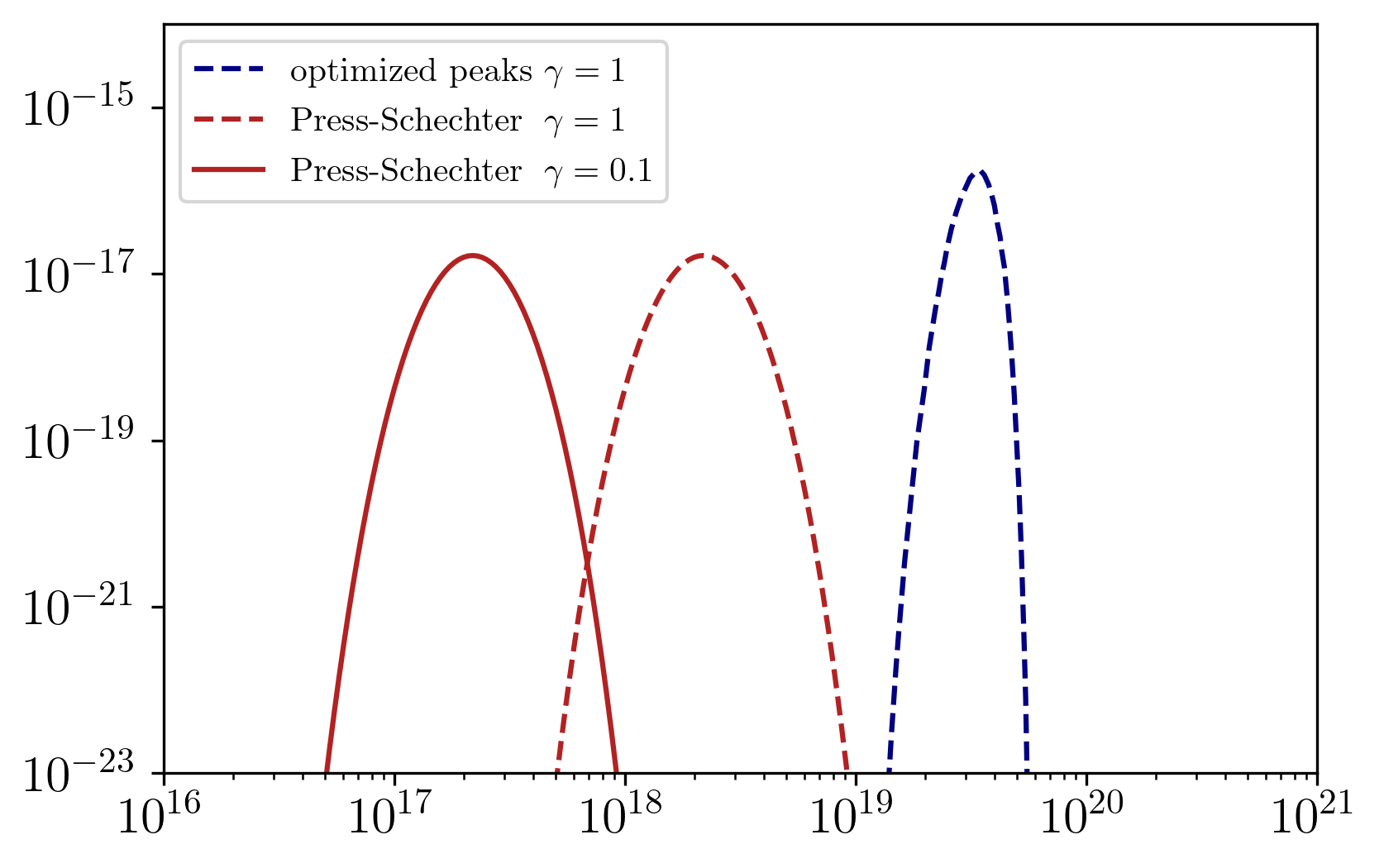}
\includegraphics[scale=0.55]{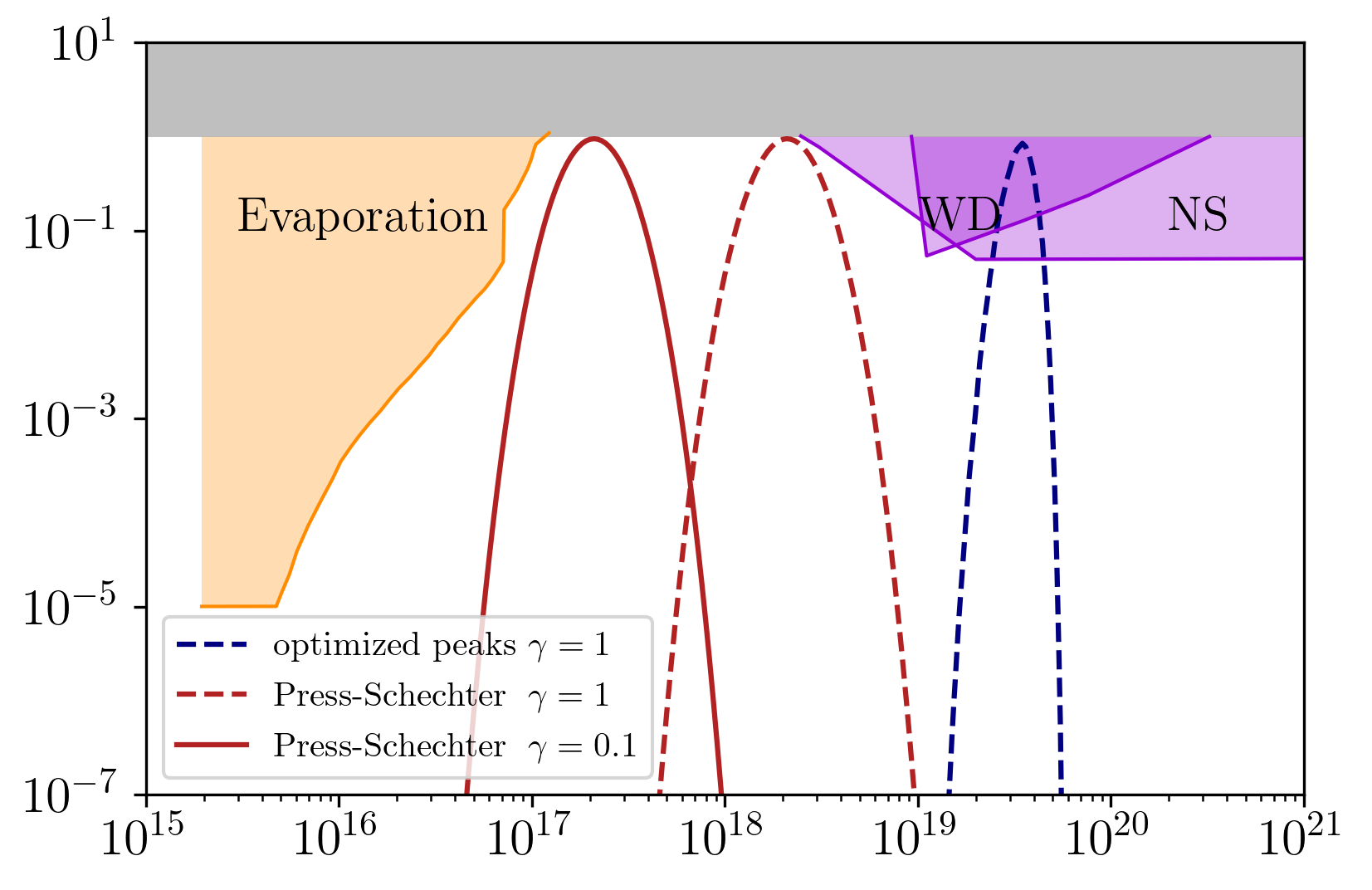}
\caption{Plots of the PBH abundance $\beta$ at the time of formation (left panel) and $f_{\text{PBH}}(M)$ (right panel), which is proportional to $\beta$ at the time of matter-radiation equality. The factor, $\gamma$, refers to the fraction of the horizon patch that goes into PBH. The shaded regions illustrate the various PBH CDM constraints in the mass range obtained from \cite{bradkav}.}
\label{fig:beta}
\end{figure*}

Equation \eqref{eq:beta1} can be simplified by introducing dimensionless quantities and other subsitutions. For example, the combination $\sigma_{2}^{2}/(\sigma_{0}\sigma_{1}^{3})=(\mathcal{F}_{2}/\mathcal{F}_{1}^{3/2})(k_{0}/\sigma_{0}^{2})$ (here $\mathcal{F}(n,\Delta)$ has been abbreviated as $\mathcal{F}_{n}$). Also, labelling $k_{0}^{2}M/(k_{\text{eq}}^{2}M_{\text{eq}})=\widetilde{M}$ and $k_{*}/k_{0}=x$ the expression for $\beta$ then becomes
\begin{align}
\beta_{\text{pk}}(M)&\approx 2\gamma\left( \frac{3}{2\pi} \right)^{1/2}\frac{\mathcal{F}_{2}}{\mathcal{F}_{1}^{3/2}}\widetilde{M}^{3/2}x\frac{\tilde{\sigma}^{2}(k_{*})}{\sigma_{0}^2}\nonumber\\
&\times f\left( \frac{\mu_{\text{b}} k_{*}^2}{\sigma_{2}} \right)P_{1}\left( \frac{\mu_{\text{b}}}{\sigma_{0}},\frac{\mu_{\text{b}} k_{*}^2}{\sigma_{2}} \right)\nonumber\\
&\times \bigg\lvert\frac{\dd \ln k_{0}r_{m}}{\dd x}-\mu_{\text{b}}\frac{\dd g_{m}}{\dd x} \bigg\lvert^{-1} 
\end{align}
The quantities $\mathcal{F}_{1}$ and $\mathcal{F}_{2}$ can be obtained from Table (\ref{table:table1}) and $M_{\text{eq}}\approx 7\times 10^{50}\text{g}$ and $k_{\text{eq}}\approx 0.01\text{Mpc}^{-1}$ \cite{Green:2004wb}. The dependence of $\beta_{\text{pk}}$ on the term in the modulus is mild and one can show that $|\frac{\dd \ln k_{0}r_{m}}{\dd x}-\mu_{\text{b}}\frac{\dd g_{m}}{\dd x} |^{-1}\sim\mathcal{O}(1)$. Equation \eqref{eq:beta1} can be constrasted with the simple Press-Schechter computation of the PBH abundance. Here, the overdensities $\delta(t,\bm{x})$ are usually considered in the linear approximation of Eq. \eqref{eq:density} such that, in Fourier space, they can be written as 
\begin{equation}
\delta(t,\bm{k})=-\frac{2(1+w)}{5+3w}\left( \frac{k}{aH} \right)^{2}\zeta(t,\bm{k})
\end{equation}
The overdensities are smoothed by a suitably chosen window function $W(kR)$ over the lengthscale defined by $R$. A popular choice for the window function is a Gaussian filter of the form $W(kR)=e^{-k^2 R^2 /2}$. With these window functions there are associated mass scales $M=\gamma_{W}\bar{\rho}R^{3}$ where $\gamma_{W}$ is a constant factor which depends on the type of window function chosen. Using the window function, a smoothed mass variance is defined as follows
\begin{equation}
\sigma^{2}\left( M(k) \right)=\frac{16}{81}\int\dd\ln q \;W^{2}\left( \frac{q}{k} \right)\left( \frac{q}{k} \right)^{4}\mathcal{P}_{\delta}(q)
\end{equation}
The power spectrum of overdensities $\mathcal{P}_{\delta}$ shares the same definition of the curvature power spectrum. With the assumption that the overdensities have a Gaussian distribution of the form
\begin{equation}\label{eq:PS}
P\left( \delta \right)=\frac{1}{\sqrt{2\pi\sigma^{2}(M)}}e^{-\delta^{2}/2\sigma^{2}(M)}
\end{equation}
the ansatz of the Press-Schechter formalism states that the fraction of collapsed objects with mass greater than $M$ is computed by integrating Eq. \eqref{eq:PS} above the threshold density for collapse $\delta_{\text{th}}$ \cite{1974ApJ...187..425P}. The PBH abundance computed this way is expressed as
\begin{equation}
\beta_{\text{PS}}(M)=\int_{\delta_{\text{th}}}\frac{\dd\delta}{\sqrt{2\pi\sigma^{2}(M)}}e^{-\delta^{2}/2\sigma^{2}(M)}
\end{equation}
Regardless of the method used to compute $\beta(M)$, since $\beta\equiv \rho_{\text{PBH}}/\rho_{\text{rad}}$, the formation fraction grows proportional to $a$ up until matter-radiation equality which then sets the fraction of PBHs that can constitute CDM. If $\beta^{\text{form}}$ denotes the abundance at formation, at equality it evolves into the following
\begin{equation}
\beta^{\text{eq}}(M)=\left( \frac{M_{\text{eq}}}{M} \right)^{1/2}\beta^{\text{form}}(M)=\widetilde{M}^{-1/2}\frac{k_{0}}{k_{\text{eq}}}\beta^{\text{form}}(M)
\end{equation}
Using this, the contribution of PBHs to the matter content of the Universe is computed
\begin{equation}\label{eq:Omega_PBH}
\Omega_{\text{PBH}}=\int_{M^{*}}^{M_{\text{eq}}}\dd\ln M\beta^{\text{eq}}(M)
\end{equation}
where the lower limit $M^{*}\simeq 3\times 10^{12}\text{g}$ corresponds to those PBHs which are massive enough to survive till matter-radiation equality before evaporating \cite{Garcia-Bellido:2017mdw}. Another important quantity is the fractional abundance of PBH over CDM, $f_{\text{PBH}}(M)\equiv \Omega_{\text{PBH}}/\Omega_{\text{CDM}}\approx 2.4\beta^{\text{eq}}(M)$, where the most recently measured value of the CDM relic abundance is $\Omega_{\text{PBH}}=0.26$ \cite{Aghanim:2018eyx}. The PBH abundances, both at formation and matter-radiation equality, have been plotted in Fig. (\ref{fig:beta}) where the red and blue curves refer to the Press-Schechter and optimized peaks calculations respectively. The dotted lines have been used to distinguish results obtained using different values of $\gamma$. In Eq. \eqref{eq:PBH_mass2}, this factor was left unspecified and indeed there are some doubts as to what this factor should be. The plots have been made for $\gamma=0.1$ and $\gamma=1$ which are the values used in Ref. \cite{Mahbub:2019uhl} and \cite{Yoo:2018kvb} respectively. The shaded regions illustrate constraints on PBH CDM in the stated mass range: (i) constraints arising from various observations related to PBH evaporation (orange) and (ii) white dwarf/neutron star populations (violet). Details on these can be found in \cite{Carr:2020gox,Green:2020jor}. For the calculation, $k_{0}=1.9\times 10^{14}\;\text{Mpc}^{-1}$ and $\mathcal{P}_{\zeta}^{\text{max}}\approx 9.2\times 10^{-3}$ which correspond to a spectral tilt of $n_{s}=0.9437$. The threshold overdensities used were
\begin{itemize}
\item Press-Schechter: $\delta_{\text{th}}=0.232$
\item Optimized peaks: $\delta_{\text{th}}=0.53$
\end{itemize}
Using Press-Schechter, PBH formation at the given abundance requires a lower threshold.\footnote{The computation of $\beta$ using Press-Schechter is highly sensitive to the choice of window function. Since Gaussian fliters are adept at remove small scale features, the lower threshold reflects this feature. Results will vary if a real-space tophat filter is used, which also needs to be supplied with a suitable transfer function. See Ref. \cite{Ando:2018qdb,Tokeshi:2020tjq} for further details.} Results from numerical simulations (in the comoving gauge) tell us that gravitational collapse occurs when the compaction function reaches the threshold $\mathcal{C}_{\text{th}}\approx 0.267$ which corresponds to $\delta_{\text{th}}\approx 0.533$. The threshold used in the peaks calculation is much closer to this and it should be the case that PBHs are formed from rare and large fluctuations corresponding to the tail of the Gaussian distribution of $\zeta$ (provided there no primordial non-Gaussianities). We do observe that the $f_{\text{PBH}}(M)$ curve from the peaks calculation impinges on observational constraints. Nevertheless, this can be remedied by using a higher value of $k_{0}$ than the one quoted. For example, the collapse of perturbations of wavenumber $k_{0}\approx 1.2\times 10^{15}\;\text{Mpc}^{-1}$ produce PBHs of mass approximately $10^{15}\;\text{g}$ using the simple mass calculation using Eq. \eqref{eq:PBH_mass1}. Since the new mass calculation in Eq. \eqref{eq:PBH_mass2} predicts PBH masses to be an order of magnitude larger, the modes $k_{0}\sim 10^{15}\;\text{Mpc}^{-1}$ may produce PBHs in the unconstrained mass interval. Of course, the abundance from Press-Schechter can be applied to cases for which $\delta_{\text{th}}$ are close to the required value. However, those would require higher amplitudes for $\mathcal{P}_{\zeta}$ and, keeping in mind the future constraints from LISA in the range $10^{9}-10^{15}\;\text{Mpc}^{-1}$, it might not be a very viable choice.

\subsection{New predictions using optimized peaks theory}
\begin{figure*}[t]
\centering
\includegraphics{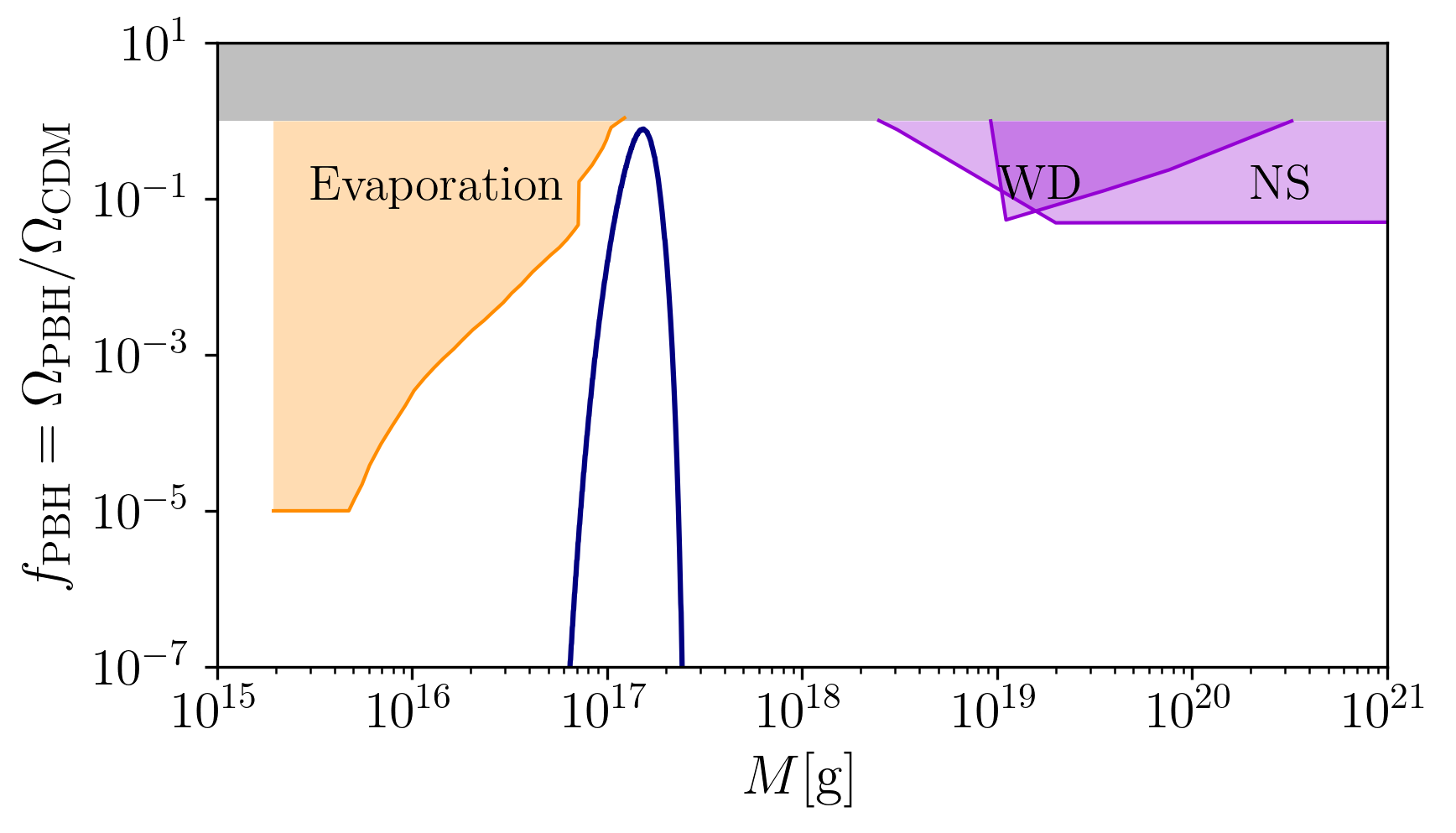}
\caption{Plot of $f_{\text{PBH}}(M)$ for $\mathcal{P}_{\zeta}(k)$ with a peak centered at $k_{0}=8\times 10^{14}\text{Mpc}^{-1}$ using optimized peaks theory. This $k_{0}$ corresponds to a spectral tilt of $n_{s}\simeq 0.9456$.}
\label{fig:beta_new}
\end{figure*}
Here the predictions from the $\alpha$-attractor model potential in Eq. \eqref{eq:potential} can be established in light of the new mass calculation given by Eq. \eqref{eq:PBH_mass2} and the techniques reviewed in Sec. (\ref{sec:PBH_abundance}). Since the mass range of $M\in\left[ 10^{17},10^{19} \right]\text{g}$ is unconstrained, with the old femtolensing constraints having been called into question, PBHs of such masses can contribute to a large portion of the observed CDM density. Exploiting the fact that there exists a $\mathcal{O}(10)$ discrepancy between the two different mass calculations the peak in the curvature power spectrum can be set for modes $k>1.9\times10^{14}\text{Mpc}^{-1}$. These typically would have produced PBHs of mass $\lesssim 10^{16}\text{g}$-- too light to be a major component of CDM and with abundance heavily restricted by evaporation constraints. Now, with the improved mass calculation, these shorter modes can be used to fill in the unconstrained mass region with a population of PBHs. In Fig. (\ref{fig:beta_new}), a more realistic plot of $f_{\text{PBH}}(M)$ is shown. This scenario can be realized by setting $k_{0}\approx 8\times 10^{14}\text{Mpc}^{-1}$ and using $\delta_{\text{th}}\approx 0.537$. These parameters result in an $f_{\text{PBH}}(M)$ peaking at around $M\approx 1.603\times 10^{17}\text{g}$-- which is close to the evaporation constraint limits, but evading them in general. The energy density of PBHs at the time of matter-radiation equality can be computed by numerically integrating Eq. \eqref{eq:Omega_PBH}, where the integration limits set by $M^{*}$ and $M_{\text{eq}}$ are not strictly necessary since the $f_{\text{PBH}}$ curve is rather narrow and sharply falls off on either side of the peak.\\
\indent Regardless of this, the numerical integration of Eq. \eqref{eq:Omega_PBH} yields $\Omega_{\text{PBH}}\approx 0.247$ and, compared to $\Omega_{\text{CDM}}=0.26$, produces $\Omega_{\text{PBH}}/\Omega_{\text{CDM}}=0.95$. This also slightly improves the value of the spectral index with $n_{s}\approx 0.9456$. With the inclusion of the running of the running, accoring to Planck 2018 \cite{Akrami:2018odb}
\begin{equation}
n_{s}=0.9587\pm 0.0112\;\;\;\;\;(95\%\;\text{C.L.})
\end{equation}
As a result, the new value of $n_{s}$ is still slightly outside the $2\sigma$ interval and the predictions are in tension with current data. Future measurements of the CMB will either validate or rule out this model as a viable candidate for producing PBH CDM in this mass range.

\section{Conclusions}
In this paper, the model proposed in \cite{Mahbub:2019uhl} has been re-examined. The model, constructed using inflationary $\alpha$-attractors, is capable of forming PBHs in abundance that can form a major component of CDM. The PBH abundance has been computed using optimized peaks theory which produces, in general, different predictions compared to the simpler Press-Schechter theory. Importantly, the peaks calculation incorporated the effects of high curvature regions on the mass of a patch of the Universe that collapses to form PBHs where there is a multiplicative correction of the form $e^{-\mu g_{m}}$. This subsequently increases the mass contained within the collapsed region, making PBHs more massive. Using these, it has been seen that PBHs can be formed in sufficient abundance using numerically favored threshold overdensities ($\delta_{\text{th}}\sim 0.5$) and PBHs can be formed in the unconstrained mass range of $10^{17}$ to $10^{19}\text{g}$. In particular, due to the mass correction introduced by effects of curvature perturbations, a $\mathcal{P}_{\zeta}(k)$ peaked at $k_{0}=8\times 10^{14}\text{Mpc}^{-1}$ can be used which previously would have predicted the formation of much lighter PBHs. As a result, the model is able to account for the totality of CDM (95\% shown in this paper) using numerically favored $\delta_{\text{th}}$ and lower amplitude peaks in the curvature power spectrum. The latter can help evade future GW constraints on the power spectrum. Lower amplitudes of $\mathcal{P}_{\zeta}(k)$ can certainly be explored- although such cases would necessarily require $\delta_{\text{th}}<0.5$.\\
\indent It would not have been unreasonable to think that this would drastically improve the $n_{s}$ value of this particular model: especially since the peaks in $\mathcal{P}_{\zeta}(k)$ could be placed at larger values of $k$. However, as seen from Fig. (\ref{fig:power_spectrum}), the increase in $n_{s}$ is only slight and does not place it within $2\sigma$ of the Planck 2018 measured value of $n_{s}$ with the running of the running considered. 

\section{Acknowledgements}
The author thanks the anonymous referee for pointing out certain inconsistencies in the original manuscript.

\bibliography{PBH_peaks}

\end{document}